\documentclass[a4paper,twoside]{article}

\usepackage{epsfig}
\usepackage{subcaption}
\usepackage{calc}
\usepackage{amssymb}
\usepackage{amstext}
\usepackage{amsmath}
\usepackage{amsthm}
\usepackage{multicol}
\usepackage{pslatex}
\usepackage{apalike}
\usepackage{float}

\usepackage{todonotes}
\newcommand{\todoi}[1]{\todo[inline]{#1}}

\usepackage[utf8]{inputenc}
\usepackage[english]{babel}
\usepackage[autostyle]{csquotes}
\usepackage[multidot]{grffile} %
\usepackage{nth}
\usepackage{paralist}
\usepackage{url}
\usepackage{comment}

\usepackage{siunitx}

\usepackage{enumitem}
\usepackage{listings}
\usepackage{makecell}
\usepackage{booktabs}
\usepackage{xspace}
\usepackage{float}
\usepackage{textcomp}

\usepackage{graphicx}
\usepackage{amsmath}

\usepackage[hidelinks]{hyperref}
\usepackage{cleveref}
\usepackage{academicons}

\floatstyle{boxed}
\newfloat{prot}{!h}{prf}
\floatname{prot}{Protocol}
\Crefname{prot}{Protocol}{Protocols}

\floatstyle{boxed}
\newfloat{scheme}{ht}{scf}
\floatname{scheme}{Scheme}
\Crefname{scheme}{Scheme}{Schemes}

\newfloat{lstfloat}{ht}{lstf}
\floatname{lstfloat}{Listing}
\Crefname{lstfloat}{Listing}{Listings}

\newfloat{tfloat}{ht}{lstf}
\floatname{tfloat}{Table}
\Crefname{tfloat}{Table}{Tables}

\setlist[description]{leftmargin=0.1in}
\setlist[itemize]{leftmargin=0.15in}

\newcommand{\mathcmd}[1]{\ensuremath{#1}\xspace}

\newcommand{\rom}[1]{(\uppercase\expandafter{\romannumeral #1\relax})}
\newcommand{\on}[1]{\textbf{on #1}}

\usepackage{SCITEPRESS}     %

\begin{document}

\title{Offline-verifiable Data from Distributed Ledger-based Registries}

\author{\vspace{-2em}
\authorname{Stefan More\sup{1}\orcidAuthor{0000-0001-7076-7563}, Jakob Heher\sup{1}, Clemens Walluschek\sup{1}}
\affiliation{\sup{1}Institute of Applied Information Processing and Communications (IAIK),\\ Graz University of Technology, Graz, Austria}
\email{\{stefan.more, jakob.heher\}@iaik.tugraz.at, official.avocari@gmail.com}
\vspace{-3em}
}

\keywords{\scriptsize This is the full version of a paper that was presented at SECRYPT 2022. \href{https://doi.org/10.5220/0011327600003283}{DOI: 10.5220/0011327600003283}.}

 \abstract{
  Trust management systems often use registries to authenticate data, or form trust decisions.
  Examples are revocation registries and trust status lists.
  By introducing distributed ledgers (DLs), it is also possible to create decentralized registries.
  A verifier then queries a node of the respective ledger, e.g., to retrieve trust status information during the verification of a credential.
  While this ensures trustworthy information, the process requires the verifier to be online and the ledger node available.
  Additionally, the connection from the verifier to the registry poses a privacy issue, as it leaks information about the user's behavior.
  In this paper, we resolve these issues by extending existing ledger APIs to support results that are trustworthy even in an offline setting.
  We do this by introducing attestations of the ledger's state, issued by ledger nodes, aggregatable into a collective attestation by all nodes.
  This attestation enables a user to prove the provenance of DL-based data to an offline verifier.
  Our approach is generic.
  So once deployed it serves as a basis for any use case with an offline verifier. %
  We also provide an implementation for the Ethereum stack and evaluate it,
  demonstrating the practicability of our approach.
}

\onecolumn \maketitle \normalsize \setcounter{footnote}{0} \vfill

\section{I{\normalsize NTRODUCTION}}

A trust management system answers trust questions by executing a policy defined by its operator.
Such a policy specifies what credentials a user needs to provide to access a resource, and which rules a credential needs to fulfill to be accepted as ``trusted'' by the system.
The policy is evaluated with regard to a set of trust information coming from different sources.
On one side, the \textit{user} (prover) provides their own credentials as input to the system.
On the other side, the \textit{verifier} often requires more information to authenticate those credentials, such as their revocation status or the trustworthiness of their issuer~\cite{DBLP:conf/openidentity/AlberMMS21,DBLP:conf/ifiptm/ModersheimSWMA19}.
Since the verifier needs to trust this additional information, it is usually collected directly from respective authorities and registries.

A distributed ledger (DL) is an attractive technology to maintain such registries.
A classic example is the common use case of revocation: Since credentials expire and mistakes happen, issuers want to be able to revoke a credential.
In the Self-Sovereign Identity (SSI) world, this is often realized using a revocation registry in form of a list stored on a DL\@.
Other examples are projects that use a DL to establish a Web of Trust as a distributed trust store and storage for credential schema information~\cite{FutureTrustConsortium2017,DBLP:conf/sec/MoreGHAK21}.

\paragraph{Challenge: Availability \& Privacy}

To retrieve DL-based data, the verifier communicates with the API of a DL node it trusts.
While this ensures the freshness of the data, a network connection to this node is required.
If the verifier is offline, it cannot retrieve a trustworthy copy of the data~\cite{DBLP:conf/trustcom/AbrahamMRH20}.
The same is true if the particular DL node used by the verifier is unavailable~\cite{DBLP:conf/ndss/LiCLT0L21}.

User privacy poses an additional challenge.
Such approaches don't provide \textit{unobservability} of interactions -- in other words, the contacted DL node learns about the showing of a credential, and about which verifier the credential was shown to.
Since verifiers are typically operated by the individual service provider, this correlates with the user's associations~\cite{DBLP:conf/imc/ChungL0CLMMRSW18}.
Often, sensitive information such as physical location can be derived.

As data provided by the DL API is currently not~signed, trust in it is derived solely from the authenticity of the underlying connection with a trusted node.
This is in contrast with comparable technologies, such as OCSP stapling in TLS~\cite{RFC:OCSPstapling}.

\paragraph{Contribution}

In this paper, we %
solve the described problem with a generic \textit{Ledger State Attestation (LSA)} system (cf.\ \Cref{sec:approach}).
Using this LSA system, a user can retrieve data from the DL and prove its provenance to an offline verifier.
Since our approach provides a generic interface to the data stored on the ledger, the system can be used in different use cases.

\paragraph{\rom{1} Node Attestations:}
We enable DL nodes to issue signed \emph{node attestations} to users (cf.\ \Cref{sec:la}).
Such an attestation contains the result of some specified operation on the DL, such as retrieving the current block hash or the result of a smart contract invocation.
Additionally, it attests in an offline-verifiable way that the result matches the node's current view of the DL state.
We achieve this without modifications to the code of the ledger clients but instead provide a wrapper around the node API\@.
This approach is also transparent to the consensus protocol used by the ledger.
The wrapper provides a generic attestation functionality and thereby supports all kinds of current and future use cases.
Although this wrapper needs to be hosted directly on the nodes' servers, this only needs to be done once.

\paragraph{\rom{2} Aggregate Attestations:}
We enable an user to retrieve such node attestations from multiple nodes aggregated into a single \emph{aggregate attestation} (cf.\ \Cref{sec:la}).
By retrieving node attestations from an appropriate set of DL nodes, the user can be reasonably sure that the aggregate attestation also includes node(s) that an unknown verifier trusts.
The verifier can then verify the attestation without needing to communicate with the node(s) in question.
As it can now trust the provided result, it can then use it to authenticate the user's credentials while remaining fully offline and without leaking information to the node or other third parties.

\paragraph{\rom{3} Implementations:}
We demonstrate the feasibility of our approach using two implementations (cf.\ \Cref{sec:implementation}).
While our general architecture is independent of a concrete DL technology, in this proof of concept implementation we focus on the Ethereum stack.
Our first variant enables DL nodes to attest the current block hash, which then allows an offline verifier to %
establish trust in any DL-based data that the user provides.
The second variant issues attestations of returned data from smart contract function calls.
This enables users to specify custom queries or filters for the data they want to retrieve.

\subsection{R{\normalsize ELATED} W{\normalsize ORK}}
\label{sec:relatedwork}

Various systems and methods in the blockchain world use data stored in a DL, but all of those systems have online components that directly interact with the ledger.
E.g., Layer 2 protocols~\cite{DBLP:conf/fc/GudgeonMRMG20} move a large number of transactions from a ledger to an off-chain service to increase performance and reduce cost.
Systems based on the layer 2 approach interact with a smart contract and thus require a connection to the DL\@.
The same requirement exists for Ethereum's Light Client, which fetches a state root from a trusted node~\cite{DBLP:journals/iacr/ChatzigiannisBC21a}.
Another example is inter-ledger communication~\cite{DBLP:conf/fc/ZamyatinAZKMKK21}, which is used to transfer assets from one ledger to another.
This transfer requires a trusted third party with a connection to both ledgers.

To prove the provenance of data, TLS-N\footnote{\url{https://github.com/tls-n}} uses a more generic approach by extending the TLS handshake to enable a server to notarize a TLS session. %
\mbox{TLSNotary}~\cite{TLSnotaryA} and \mbox{DECO}~\cite{DBLP:conf/ccs/ZhangMMGJ20} are concerned with the attestation of %
access protected web data to a third party.
This is realized by involving a third party (oracle) trusted by the verifier in the TLS session with the server.

Some revocation systems work without a direct connection between the verifier and the revocation authority.
One common example of this is the verification of TLS certificates:
OCSP stapling is a TLS extension that allows a certificate subject (web server) itself to acquire the status information of its certificate~\cite{RFC:OCSPstapling}.
This information is signed by a status authority, which ensures that a verifier can trust the information.
The subject can then provide it to the verifier (web browser), and the verifier does not require a direct connection to the status authority.

The system by Abraham et al.~\cite{DBLP:conf/trustcom/AbrahamMRH20} allows the offline verification of DL-based revocation information.
However, it is limited to only the revocation use case.
Modifying this system for other use cases is possible, but each additional use case requires adaptions on all of the DL's nodes.

\subsection{B{\normalsize ACKGROUND}}
\label{sec:background}

\paragraph{Distributed Ledger (DL)}

A DL is a redundant append-only datastore on distributed nodes without central control~\cite{DBLP:conf/middleware/JannesLJ19,DBLP:conf/trustcom/AlexopoulosDMH17}.
The nodes agree on the ledger's current state by running a consensus protocol~\cite{DBLP:journals/comsur/XiaoZLH20}.
The distribution of nodes can be geographical, political, institutional, etc.\ to prevent collusion and improve resilience.
DLs can be classified into different access models, depending on who can join the network
(public vs.\ private), or by who has read and write access to it (permissioned vs.\ permissionless).
In this work we focus on permissioned ledgers.

\paragraph{Smart Contract (SC)}

Many ledgers support the storing of code on the DL, which is then deterministically executed by the nodes performing the consensus protocol.

Examples of this are the Ethereum ledger\footnote{\url{https://docs.soliditylang.org}} and various ledgers from the Hyperledger project.\footnote{\url{https://hyperledger-fabric.readthedocs.io/en/release-2.2/chaincode4ade.html}}
One particular example is Hyperledger Besu,\footnote{\url{https://besu.hyperledger.org}} a Ethereum client specifically designed for use in permissioned consortium ledgers.
Such code is called a ``smart contract''~\cite{buterin2014next} in the Ethereum world, while the Hyperledger project also calls it ``chaincode''~\cite{cachin2016architecture}.
Variables in the code are stored on the DL as well and can be read and modified using functions supplied by the contract.
SC code is written in a high-level language and then compiled to ledger-specific bytecode. Only this bytecode is then written to the DL\@.
When a user sends a function call to a contract, nodes execute this bytecode, for example using the Ethereum Virtual Machine (EVM).\footnote{\url{https://ethereum.org/en/developers/docs/evm}}
The resulting state is only written to the ledger if all nodes agree on the result of the computation.

SCs can be used to provide simplified views on complex data stored on the DL, %
forming a generic query and filtering system, akin to stored procedures in a traditional database.

\subsubsection{Ethereum JSON-RPC API \& web3.js}

To allow other entities to access the state of the DL and call SC functions, Ethereum nodes provide an HTTP API\@.\footnote{\url{https://eth.wiki/json-rpc/API}}
It offers a JSON-RPC interface, which can be used by entities who do not wish to operate a full node, or are not able to participate in the ledger.

On the client-side the web3.js\footnote{\url{https://github.com/ethereum/web3.js}} library is commonly used to interact with nodes' JSON-RPC API\@.
It provides users a high-level interface to interact with SC functions and translates these function calls to a representation that the EVM understands.

\paragraph{BLS (Multi-)Signatures}

Boneh–Lynn–Shacham (BLS) is a provable secure pairing-based signature scheme for producing short signatures~\cite{DBLP:journals/joc/BonehLS04}.
One property of  BLS is that it can be used as a multi-signature scheme; signatures by multiple private keys can be combined into a single constant-size aggregate signature.
This saves space and verification time~\cite{DBLP:conf/pkc/Boldyreva03,DBLP:conf/asiacrypt/BonehDN18}.

\section{D{\normalsize ESIGN}}
\label{sec:design}
\label{sec:approach}

\begin{figure}[h]
    \centering
    \includegraphics[width=1\linewidth]{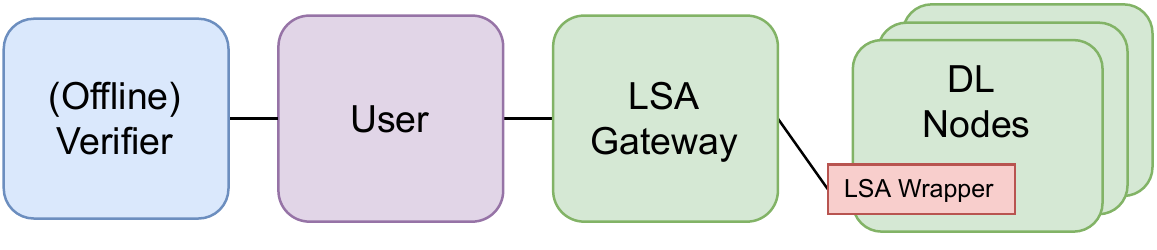}
    \caption{High-level architecture of our Ledger State Attestation system.}\label{fig:laarch}
\end{figure}

\begin{figure*}[h]
    \centering
    \includegraphics[width=\linewidth]{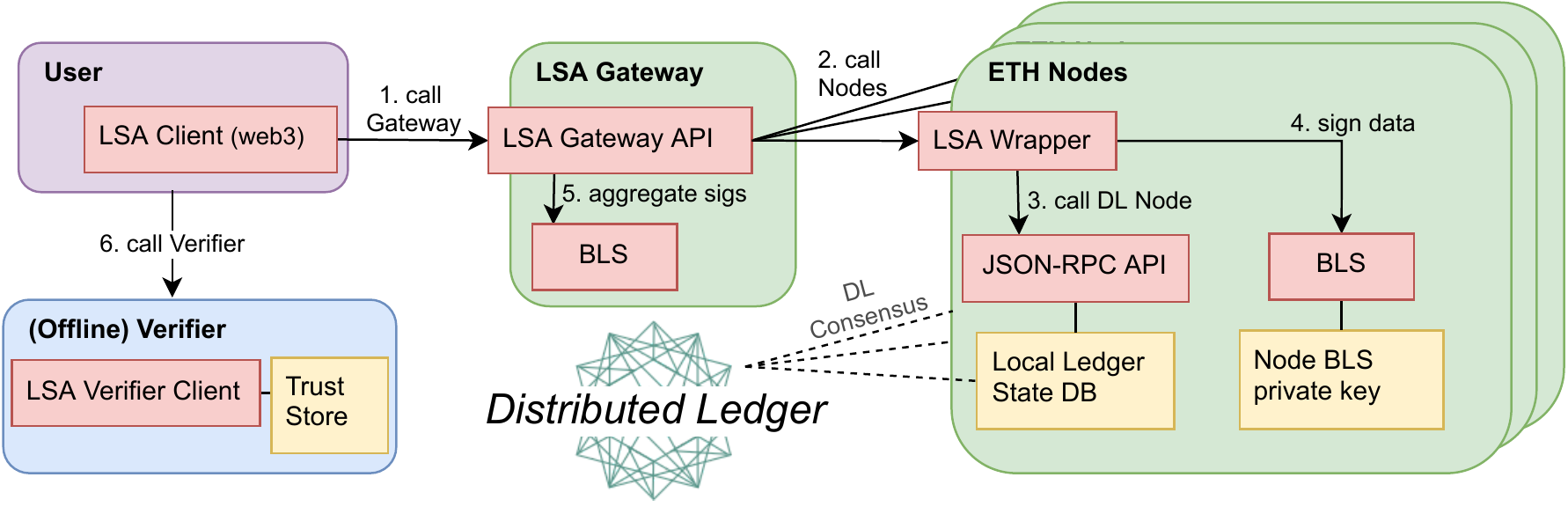}
    \caption{The architecture of our Ledger State Attestation system, extending the functionality of Ethereum nodes and web3.js.}\label{fig:laflow}
\end{figure*}

There are several main components in our system, which we describe below.
A high-level and generic overview of these components and how they are connected is shown in \Cref{fig:laarch}, while \Cref{fig:laflow} offers a more detailed and Ethereum-focused picture.

The distributed ledger (DL) is represented by its \textbf{DL Nodes}.
These nodes communicate peer-to-peer, using a consensus protocol to agree on a shared state.
DL Nodes provide an HTTP API to allow other entities to access the state of the DL\@.

We add the \textbf{LSA Wrapper} component to each DL node.
It wraps the node API, providing access to the same data but enriches the API functionality, adding a proof of provenance to the returned data.
The wrapper provides the same API endpoints as the node API, so it is compatible with existing API clients.

We introduce the \textbf{LSA Gateway} stand-alone component to support the user by retrieving data from all DL nodes.
It can be part of a node, run by the user, or be operated by a third party.
We discuss the implications of this choice in more detail in \Cref{sec:discussion}.
The gateway provides the same API endpoints as the node API and our LSA Wrapper.
It forwards queries it retrieves from a user to the DL nodes, and aggregates their answers into a single response for the user.

The \textbf{User} wants to retrieve data from a DL, and present it to a verifier.
To retrieve data directly from the DL, the users interact with the API of a DL node.
For data that can be presented to an offline verifier later, they instead interact with the LSA Gateway.

The \textbf{(Offline) Verifier} receives data from the user and needs to establish trust in this data.
We consider a scenario where this verifier is offline, i.e., it cannot connect to any of the DL's nodes during verification.

In DL-based trust management, the verifier is interested in the trustworthiness of a \textbf{claim} about the DL's state.
This claim can, e.g., be an assertion of the current block hash, or of the return value of a smart contract function.
We define an \textbf{attestation} as data combined with proof of authenticity.
There are two types of attestations:
A \textbf{node attestation}, created by an individual DL node, attests that the data reflects the data in its local storage.
Combining such attestations of several nodes yields an \textbf{aggregate attestation}, attesting an agreement between all involved nodes. %
If an aggregate attestation contains the attestations of \textit{all} nodes of a DL, we call this a \textbf{ledger attestation}.

\subsection{Ledger State Attestation (LSA)}
\label{sec:la}

When a user shows some data to a verifier, this verifier needs to make sure that it can trust this data before relying on it for further processing.
Trusting data coming from a DL-based registry means verifying that this data matches the data stored in the DL\@.
An online verifier could check this by contacting a trustworthy DL node and comparing the data, or even by doing additional lookups on its own.

Since we consider the scenario where a verifier is offline, this online check is not possible.
The underlying challenge is that an offline verifier has no reason to trust data that a user claims to have retrieved directly from a node's API\@.
For example, a verifier cannot be sure if this data was really retrieved from a DL, that the user did not alter the data, or that the data represents the latest state of the DL\@.

\paragraph{Attestation of state by a node}

To mitigate this problem, we add the LSA Wrapper component to all nodes of the DL, wrapping the node API\@.
This is the only modification to a DL node our approach requires, and it only needs to be done once and not for every use case.
The wrapper enables DL nodes to issue attestations of data stored on the DL\@:
While the default node API answers user queries by returning plain data, the wrapper additionally adds a proof to the response.
To ensure the authenticity of the data, the wrapper creates this attestation proof using a private key of the node.
To enable a verifier to decide if the presented information was fresh enough, the number of the current block and a low-resolution timestamp are also added to the attestation.

The attestation can then later be presented to an offline verifier which checks it to ensure that the presented data was returned by a specific DL node and has not been altered.
After receiving some attestation (data and proof) from a user, the verifier uses the node's public key from a local trust store to verify the attestation, before processing the data.

\paragraph{Attestation by the whole DL}

While this process ensures authenticity (and integrity) of the data with respect to one node, it means the user needs to select a node the verifier trusts.
This is a problem since a user does not know, at the time of retrieving the attestation, which node(s) a verifier trusts.
Additionally, this limits which verifier the user can present an attestation to, since different verifiers trust different nodes.
To avoid this, the user would need to retrieve an attestation by all DL nodes.

In our LSA system, the gateway is used to provide users with an easy way to retrieve data, alongside a proof from all nodes.
Since the data stored by each node was agreed upon using the consensus protocol, the data returned is also the same for each node.
But the proof returned by the nodes is different since it proves authenticity for a certain node.
This allows that the gateway returns the data only once, and aggregates the proofs of the nodes into a single proof.

\newcommand*{\sk}{\mathsf{sk}} %
\newcommand*{\pk}{\mathsf{pk}} %
\newcommand{\SIG}{\mathcmd{\mathsf{SIG}}}
\newcommand{\MSIG}{\mathcmd{\mathsf{MSIG}}}
\newcommand{\Sign}{\mathcmd{\mathsf{Sign}}}
\newcommand{\Verify}{\mathcmd{\mathsf{Verify}}}
\newcommand{\ASigs}{\mathcmd{\mathsf{ASigs}}}
\newcommand{\Encode}{\mathcmd{\mathsf{Encode}}}

\newcommand{\nodeclient}{\mathcmd{\mathsf{Client}}}
\newcommand{\node}{\mathcmd{\mathsf{Node}}}
\newcommand{\gateway}{\mathcmd{\mathsf{Gateway}}}
\newcommand{\attest}{\mathcmd{\mathsf{Attest}}}
\newcommand{\execute}{\mathcmd{\mathsf{Execute}}}
\newcommand{\query}{\mathcmd{\mathsf{query}}}
\newcommand{\call}{\mathcmd{\mathsf{call}}}
\newcommand{\parameters}{\mathcmd{\mathsf{parameters}}}
\newcommand{\data}{\mathcmd{\mathsf{data}}}
\newcommand{\epoch}{\mathcmd{\mathsf{epoch}}}
\newcommand{\timestamp}{\mathcmd{\mathsf{low{\text -}resolution~timestamp}}}
\newcommand{\request}{\mathcmd{\mathsf{request}}}
\newcommand{\dlnode}{\mathcmd{\mathsf{Node}}}
\newcommand{\nodeapi}{\mathcmd{\mathsf{NodeAPI}}}
\newcommand{\NA}{\mathcmd{\mathsf{NA}}} %
\newcommand{\LSA}{\mathcmd{\mathsf{LSA}}} %
\newcommand{\LA}{\LSA}

\begin{prot}[!ht]
	\begin{description}
  \item[\underline{Attestation Phase:}]~
    \begin{itemize}[label={},leftmargin=0.2cm]
     \raggedright{}  %

    \item \on{User}
			\begin{enumerate}
      \item encode the user's claim:
        $\query \gets \nodeclient'.\Encode(\call, \parameters)$
      \item send encoded $\query$ to the LSA Gateway
      \end{enumerate}

    \item \on{LSA Gateway}
			\begin{enumerate}[resume]
      \item forward $\query$ to $n$ nodes
      \end{enumerate}

    \item \on{Node $i$ of the DL (in parallel)}
			\begin{enumerate}[resume]
      \item retrieve data by executing the user's $\query$ on the DL's state:
        $\data \gets \dlnode.\execute(\call, \parameters)$
      \item create a node attestation statement:
        $\epoch \gets \timestamp$
        $\NA_{i}' \gets (\call, \parameters, \data, \epoch)$
      \item create a multi-signature:
        $\sigma_{i} = \MSIG.\Sign(\NA_{i}', \sk_{i})$
      \item issue node attestation $\NA$ to the LSA Gateway:
        $\NA_{i} \gets (\NA_{i}', \sigma_{i}, \pk_{i})$
      \end{enumerate}

    \item \on{LSA Gateway}
			\begin{enumerate}[resume]
      \item receive node attestations from $n$ nodes and aggregate them:
        $\sigma = \MSIG.\ASigs(\{\sigma_{i}\}_{\forall i \in [n]})$
      \item issue ledger attestation $\LA$ to the user:
        $\LA \gets (\call, \parameters, \data, \epoch, \sigma,
        \{\pk_{i}\}_{\forall i \in [n]})$
      \end{enumerate}

    \item \on{User}
			\begin{enumerate}[resume]
      \item receive $\LA$ and store it for the specified $\call$ and
        $\parameters$
      \item (\textit{Optional:} verify $\LA$ using local trust store)
      \end{enumerate}

		\end{itemize}
	\end{description}
	\caption{Attestation Protocol}
	\label{prot:attestation}
\end{prot}

\begin{prot}[!ht]
	\begin{description}
  \item[\underline{Showing Phase (Offline):}]~
    \begin{itemize}[label={},leftmargin=0.2cm]
      \raggedright{}  %

    \item \on{User}
			\begin{enumerate}
      \item receive the verifier's request for DL data, which specifies the required $\call_{V}$ and the values for some
        $\parameters_{V}$ while leaving the values of other parameters $\parameters_{U}$ to the user
      \item retrieve stored $\LSA$ for the specified $\call$ and send it to the verifier
      \end{enumerate}

    \item \on{Verifier}
			\begin{enumerate}[resume]
      \item verify the attestation $\LA = (\call, \parameters, \data, \epoch, \sigma,
        \{\pk_{i}\}_{\forall i \in [n]})$:
        \begin{itemize}
        \item $\MSIG.\Verify(\LA) = 1$
        \item check if enough signer keys $\{\pk_{i}\}_{\forall i \in [n]}$ are part of the trust store
                  and that $\epoch$ is fresh enough
        \item verify that $\call = \call_{V}$ and $\parameters = (\parameters_{V}, \parameters_{U})$
        \item check if $\parameters_{U}$ and $\data$ fulfill the verifier's policy
        \end{itemize}
      \item use the now-trusted $\data$ for further processing
      \end{enumerate}

		\end{itemize}
	\end{description}
	\caption{Offline Showing Protocol}
	\label{prot:showing}
\end{prot}

\Cref{prot:attestation} shows all the required steps in this attestation process, while \Cref{prot:showing} does the same for an interactive showing to an offline verifier.
Our protocol relies on a multi-signature scheme $\MSIG$ for the attestation proof.

\section{I{\normalsize MPLEMENTATION}}
\label{sec:implementation}

To show the feasibility of our LSA concept, we implement a prototype for the Ethereum stack.
In this section and the following subsections, we describe how we applied our approach to the API of Ethereum nodes, and how the implementation can be used in practice. %
We focus on permissioned DLs, such as those used in the European Blockchain Services Infrastructure (EBSI)~\cite{EBSI}.

On the server-side, we provide a wrapper for Ethereum's RPC API, which extends the API of Ethereum nodes to support signed responses.
The LSA Gateway first forwards the user's query to the wrapped API of all applicable nodes.
Then, it aggregates the received node attestations into one aggregate attestation.
For the authenticity proofs, we use digital signatures issued individually by each node using their private key.
To be able to aggregate the signature of the individual nodes into one combined signature, we use the BLS signature scheme for the authenticity proofs~\cite{RFC:BLS}.
An additional benefit of using BLS is its efficiency and small storage requirements, minimizing the overhead (cf.\ \Cref{sec:discussion:evaluation}). %

On the client side, we extend the \emph{web3.js} library\footnote{\url{https://github.com/ethereum/web3.js}} to retrieve node and aggregate attestations. %
This allows a user to call contracts using well-established high-level calls
, but retrieve the response value in a signed and offline-verifiable form.

To demonstrate the flexibility of our design, we implement two different variants, which differ in the type of data that gets attested.

\paragraph{Variant 1: Attestation of Data} %

In this variant %
we enable an offline verifier to trust any raw data retrieved from the DL by the user.
This then allows the verifier to execute a locally stored smart contract, or work with the data by other means.

Ethereum ledgers protect the integrity of their data using merkle trees:
The block hash is the root hash of a merkle tree, formed by all transactions, smart contract code, and data stored in the ledger at a certain point in time~\cite{EthereumYellowPaper}.
A trustworthy attestation of the block hash therefore allows any data in that block, and any previous block, to be trusted.
Another advantage of the attestation of the root hash is that it can be pre-computed.
Since this needs to be done only once per block, this significantly reduces the load on the DL nodes while still allowing to establish trust in all data on the DL\@.

Thus, we create a mechanism to retrieve an attestation of the current root hash. %
This allows a user to retrieve any raw data, and the supplementary parts of the merkle tree, called merkle proof~\cite{etgGetProof}, from any (single) node.
Afterward, they send this data, the merkle proof, and the attestation of the root hash to the verifier. %
The verifier can then use this trusted hash to establish trust in the data.

\paragraph{Variant 2: Attestation of Smart Contract Response}

In our second variant, we enable users to query for and retrieve trustworthy data from the DL\@.
To do so, we extend the functionality of the node API in a way that nodes can issue attestations for
the return value of smart contract functions.
Part of this attestation is also the address of the called smart contract, the executed function, and the call parameters.
This enables users to send queries to a smart contract and get an attestation of the query result.
To provide more flexibility, we do this by extending the \texttt{call} function of the web3 client library with the ability to request and handle signed attestations.

A user can simply execute the function in the same way as without the LSA system.
Each node then executes the function call, creates a node attestation of the return value, and sends the signature to the gateway\@, which aggregates the signatures.
The result is an aggregate attestation that contains the call and return value of a certain contract function, and proves the consensus of the nodes about this state.
This attestation credential can be shown to an offline verifier and authenticated using the verifier's truststore.
Since the attestation contains the needed data, the verifier does not need to execute a smart contract or evaluate a merkle proof.

\subsection{Attestation \&{} Showing Process}

The %
process works as follows in both variants.
The structure of this process is also shown in \Cref{fig:laflow}.

\begin{enumerate}
\item To fetch the attestation of some data, the user utilizes our modified web3 library to send the request to the LSA Gateway.
           In variant 1, this is a request for the block hash, while in variant 2 this is a call to a smart contract function including parameters.
\item The LSA Gateway has a list of all nodes of the DL\@.
          It forwards the request to all nodes, which run Ethereum's RPC API with our wrapper.
\item Each node's wrapper first forwards the request to the RPC API of the node itself, which retrieves the requested data from its local version of the ledger state.
          In variant 2, it also retrieves and executes the called smart contract function using the EVM\@.
\item After retrieving the result, each of the nodes creates a node attestation using its own BLS private key and sends the result back to the gateway.
          In variant 1, the nodes attest the retrieved root hash of the current block, while in variant 2 they attest the result of the smart contract call.
\item The LSA Gateway then aggregates all signatures and sends the aggregate attestation back to the user, who stores it, e.g., in a digital wallet.
\item Later, the user shows the aggregate attestation to an offline verifier, for example by sending the attestation by Bluetooth to the verification device.
\item The offline verifier then uses their trust store to authenticate the attestation and thereby establishes trust in the attested data.
\begin{enumerate}
    \item In variant 1, the verifier can now use the now-trusted block hash to authenticate the merkle proof the user also sent.
             It then checks the merkle proof to authenticate the rest of the data, which is only then used to locally execute a locally stored smart contract.
    \item In variant 2, the verifier checks that the smart contract address and call specification contained in the attestation are the expected values.
\end{enumerate}
\end{enumerate}

\subsection{Evaluation}
\label{sec:discussion:evaluation}

We consider the performance of our LSA approach.
To do this, we contrast it with traditional online verification.
We identify the following additional costs.

\subsubsection{Attestation Phase}

\textbf{Attestation retrieval} requires an additional network round trip compared to a traditional online query, in scenarios where the LSA Gateway is not co-located on the user device.
Quantifying this overhead exactly is difficult, as it varies based on the physical location of, and connection topology between the various entities.
However, given that even a transatlantic round trip typically takes only around 90~\si{\milli\second}~\cite{IPlatency}, we consider this to be negligible.

\textbf{Data attestation} requires each node to create a signature over the data retrieved from the DL\@.
Using BLS signatures with 128-bit security, as in our implementation, signature creation takes $\approx$0.3~\si{\milli\second} on a typical consumer laptop~\cite{RFC:BLS}.

\textbf{Data retrieval} of DL data by any one individual node incurs no additional overhead compared to the traditional online flow.
As the LSA Gateway sends queries to all nodes in parallel, the query time in the worst case should be no worse than for a single node.
However, it is worth noting that, when viewed across the entire ledger, our scheme induces additional load.
While in the traditional model the online verifier only sends its query to a single trusted node which has to perform data retrieval, in our case, the LSA Gateway sends this query to many different nodes, each of which has to perform the operation.

\subsubsection{Showing Phase}

During verification %
the user needs to \textbf{transmit the retrieved LSA} to the verifier.
Since this communication could happen on constrained devices and a slow channel, we also consider the storage size of an attestation.
In BLS, both signature and public key are encoded as single group elements~\cite{RFC:BLS}.
Thus, an aggregated signature uses 48 bytes, with an additional 48 bytes per public key.
To evaluate the impact of this storage overhead, we measure transmission of an attestation over 10~kB of data and 20 public keys using mid-range smartphones: a Samsung Galaxy XCover Pro and a Google Pixel 1.
This results in a transmission time of $\approx$150~\si{\milli\second}, even using Bluetooth 4.2.

Then, the verifier needs to \textbf{verify the LSA's signature}.
For BLS signatures with 128-bit security, this takes $\approx$2.7~\si{\milli\second} on a typical laptop~\cite{RFC:BLS}.

We note that transmission time, scaling with the size of the transmitted data and number of involved nodes, appears to be the primary driver of verification time.
This presents potential optimizations by reducing the size of the transmitted LSA\@.
For example, the public key space requirement could be removed by also aggregating the BLS public keys.
On a permissioned ledger with a stable node membership, public keys could be outright omitted from the attestation, and verification could be performed using a complete trust store located at the verifier.
Regardless, we consider a total duration overhead of $\approx$153~\si{\milli\second} to be negligible for an interactive showing~\cite{DBLP:books/crc/tucker97/Nielsen97}.

\section{D{\normalsize ISCUSSION}}
\label{sec:discussion}

\subsection{Trust Assumptions}
\label{sec:discussion:trust}

\textbf{The user} must trust the verifier's trusted DL nodes to provide truthful attestations.
This assumption is also made in the online case, and is not unique to our work.

Additionally, heading into an offline scenario the user relies on the provided attestation being valid.
It is not a negligible concern that the LSA Gateway returns a bogus attestation.
In order for the user to verify the provided attestation, they would need to have a list of all DL nodes and their keys on their device.
In general, this is not trivial (see also \Cref{sec:discussion:operational}).
Therefore, the user must trust the LSA Gateway to provide a valid attestation.
They may also retrieve attestations from multiple different LSA Gateways.
As long as at least one returns a valid attestation, the user device can successfully complete the LSA process.

\textbf{The verifier} has some trust policy that relies on the truthfulness of some subset of DL nodes.
To verify the attestation proof, the verifier also has a store with the public keys of the nodes it trusts.
This does not require additional assumptions beyond those already made in the online scenario.
The LSA Gateway simply retrieves attestations from all DL nodes, including the nodes trusted by the verifier, and aggregates them.
The verifier's trust in the aggregate attestation derives from the inclusion of attestations by nodes it trusts.
It does not need to trust the LSA Gateway.
Indeed, the existence of the gateway is transparent to the verifier.

\subsection{Operational Concerns}
\label{sec:discussion:operational}

To forward requests, the LSA Gateway requires an up-to-date list of all DL nodes.

Maintaining such a list is, in general, not a trivial task on permissionless ledgers.
It thus intuitively makes sense to separate the LSA Gateway from the user device and, for example, to include it into one (or more) well-known nodes of the DL\@.
This co-location might enable use of DL information already-available to the node to contact the other nodes.

For some DL setups like permissioned DLs, the set of nodes is well-known, static, or otherwise can be easily obtained.
In that case, it makes sense to instead include the LSA Gateway into the user's client device.
This eliminates the trust considerations towards the LSA Gateway outlined in \Cref{sec:discussion:trust}.

\paragraph{Availability:}

In practice, it may not be possible for the LSA Gateway to reach all DL nodes.
This can happen due to network issues, maintenance, or as the result of a DoS attack~\cite{DBLP:conf/ndss/LiCLT0L21}.
In this situation, it may only be possible for the LSA Gateway to provide an incomplete attestation.
At what point it should give up and do so, and how this would be communicated to the user, are open questions.

Additionally, if the verifier expects an attestation was signed by \textit{all} nodes of a certain trust subset\@, an availability issue arises:
if the LSA Gateway was not able to reach all of these nodes, the resulting incomplete attestation will be rejected.

To mitigate this, verifiers could use a threshold policy.
Using the list of public keys that are part of an attestation, the verifier first verifies the aggregated signature on the data.
It then checks if at least $k$ nodes from its trust store signed the provided aggregate, and accepts it if so.

\paragraph{Required Modifications:}

Modifications to existing systems are always a challenge, especially to nodes in a distributed system.
An advantage of our approach is that the only such modification is the addition of the LSA Wrapper to the nodes, which provides generic attestation and can thus be employed in various use cases.
Such a modification could be for example performed during the setup of the system, and only the nodes considered by any verifier need to be modified.
This is in contrast to the state of the art, where each use case requires an additional modification to the DL nodes, which is often not feasible during operation.

\subsection{Limitations \& Future Work}

\paragraph{Attestation Freshness}
A limitation of both stated approaches is the fact that information authenticated using such attestations is less up-to-date than information directly retrieved from a registry.
While this is an acceptable trade-off for some use cases, other use cases require more timely information.
Depending on the concrete requirements on freshness, we can mitigate the limitation up to some extent by utilizing the network connection of the user.
In a scenario where the user is online during or shortly before the interaction with the (offline) verifier, they can retrieve a fresh attestation.
This makes sense for a verifier operated on a constrained device and is especially useful for the user's privacy since it facilitates \textit{unobservability} of the interaction.

\paragraph{Synchronized Time-stamping}
In our scheme, we assume that the nodes queried by the LSA Gateway will typically agree on the state of the DL\@.
This results in the returned node attestations having identical content, allowing the node attestations to be aggregated into a single aggregate attestation.

However, the inclusion of a timestamp, low-resolution as it might be, in the attested claim complicates this.
For any variety of reasons, the attestations returned by two nodes may end up in two (adjacent) epochs.
Since the epoch is part of the signed data, this difference makes it impossible to aggregate the signatures.

One potential workaround would be for the LSA Gateway to include an epoch derived from its local timestamp in its query to the DL nodes.
The DL nodes could then verify that the epoch is within an acceptable interval of their local clock time, and issue their attestation with the requested epoch. %
This ensures that all nodes issue their attestations for the same epoch, thereby also for the same claim.

\paragraph{Node Discovery}
Our approach works well for permissioned DLs such as consortium ledgers, but node discovery is a challenge in open architectures.
In permissioned DLs like the European Blockchain Services Infrastructure (EBSI), the set of nodes is known and changes relatively rarely~\cite{EBSI}.
On the other hand, permissionless ledgers like mainnet Ethereum have a large and unstable set of nodes, and no node knows all other nodes.
In our naive implementation, as all nodes perform the attestation process, the LSA Gateway needs a list of those nodes.
This presents a significant challenge when applying our approach to such a permissionless ledger.

While some node discovery systems exist~\cite{DBLP:conf/iptps/MaymounkovM02},\footnote{e.g., \url{https://github.com/ethereum/devp2p/blob/master/discv4.md} and \url{https://eth.wiki/en/ideas/kademlia-peer-selection}} future work is needed to access if they are applicable for our system and in what way they can be used by a verifier to create the required trust store.

\subsection*{C{\normalsize ONCLUSIONS}}

In many previous decentralized trust systems, an implicit always-online requirement is a major hindrance to practical applicability.
We resolve this issue by applying the battle-tested concept of OCSP stapling to the distributed ledger ecosystem.

After collecting signed attestations of the ledger's current state from a sufficiently large subset of DL nodes while online, the user can present this aggregate attestation to a verifier later in an offline setting.
The verifier can use its local trust store to verify that the claimed state was attested by nodes it trusts, establishing trust in the data itself.
This allows the data to be used to make informed decisions regarding the user's credentials.

In this work, we introduced Ledger State Attestations, which allow arbitrary queries to DL nodes' HTTP API to retrieve attestated results.
This serves as the basis for almost any imaginable use case with only a single adjustment to the underlying DL's nodes, and is a significant improvement over the state of the art.
Additionally, our LSA approach enables unobservability of interactions with the verifier, which is an important property to ensure the privacy of users.

Furthermore, we provided a proof of concept implementation for Ethereum-based ledgers.
We evaluate this implementation, demonstrating the practical feasibility of our scheme.

Finally, we discuss the implications of the LSA concept in terms of performance impact, added trust requirements, and operational concerns.

\section*{A{\normalsize CKNOWLEDGEMENTS}}

This work was supported by the European Union's Horizon 2020 research and innovation programme under grant agreement \href{htps://dx.doi.org/10.3030/871473}{\textnumero~871473 (KRAKEN)}.

\vspace{-1em}

\bibliographystyle{apalike}
{\small
\bibliography{bib}}

\ifnum\thepage>10
\todoi{Page limit for all content is 10 pages! (currently \thepage)}
\fi

\end{document}